\documentstyle [twoside,12pt]{article}
\newcommand{\nc}{\newcommand}
\nc{\la}{\lambda} \nc{\La}{\Lambda}  \nc{\al}{\alpha}
\nc{\th}{\theta}  \nc{\be}{\beta}
\nc{\ga}{\gamma}  \nc{\Ga}{\Gamma}
\nc{\de}{\delta}  \nc{\De}{\Delta}
\nc{\si}{\sigma}  \nc{\ka}{\kappa}
\nc{\om}{\omega}  \nc{\Om}{\Omega}
\nc{\nf}{\infty}
\nc{\ra}{\longrightarrow}
\nc{\beq}{\begin{equation}}
\nc{\eeq}{\end{equation}}
\nc{\beqa}{\begin{eqnarray}}  \nc{\dst}{\displaystyle}
\nc{\eeqa}{\end{eqnarray}} \nc{\nnb}{\nonumber}
\title{\bf Einstein\,-Weyl structures corresponding to \\ diagonal K\"ahler
Bianchi IX metrics}
\author{Guy Bonneau\thanks
{\noindent Laboratoire de Physique Th\'eorique et des Hautes Energies,
 Unit\'e associ\'ee au CNRS URA 280,~Universit\'e Paris 7,
 2 Place Jussieu, 75251 Paris Cedex 05. bonneau@lpthe.jussieu.fr}}
\begin{document}
\maketitle
\begin{abstract}
\noindent We analyse in a systematic way the
 four dimensionnal Einstein-Weyl spaces equipped with a diagonal K\"ahler
Bianchi IX metric. In particular, we show that
the subclass of Einstein-Weyl structures with a constant conformal scalar
curvature is the one with a conformally scalar
flat - but not necessarily scalar flat - metric ; we exhibit its 3-parameter
distance and Weyl one-form. This
extends previous analysis of Pedersen, Swann and Madsen , limited to the scalar
flat, antiself-dual case. We also check that, in agreement with a theorem of
Derdzinski, the most general conformally Einstein metric in the family of
biaxial K\"ahler Bianchi IX metrics is an extremal metric of Calabi, conformal
to Carter's metric, thanks to Chave and Valent's results.
\end{abstract}

\vfill
{\bf PAR/LPTHE/96-52/hep-th/9612055}\hfill  November 1996
\newpage

\section{Introduction}

In a recent paper, Tod \cite{Todd96} exhibits the relationship between some
Einstein-Weyl geometry without torsion (the four-dimensional self-dual
Einstein-Weyl
 geometry studied by Pedersen and Swann \cite{PS93}) and some local heterotic
geometry (i.e.  the Riemannian geometry with torsion and three complex
structures, associated with  (4,0) supersymmetric non-linear $\si$ models
 \cite{{HullWitten85},{xxy},{Delduc}}).

To extend these ideas to other situations, as a first step, we analyse in this
work in a systematic way a larger class of Einstein-Weyl geometries : the ones,
recently considered by Madsen \cite{{Madsen-a},{Madsen-97}} where the metric
lies in the diagonal K\"ahler Bianchi IX family. We think that this will help
in finding new bridges with a larger class of heterotic geometries.

The paper is organised as follows : in Section 2, we describe Bianchi IX
metrics and recall the definition of Einstein-Weyl structures and the
constraints that result for the metric
 and the Weyl connection (for recent reviews see refs.
\cite{{PS93},{DS95},{Tod95},{Madsen-a}}).

The Section 3 contains the complete solution of the aforementioned constraints
and discusses some special cases previously known. We first obtain a
four-parameter Einstein-Weyl structure, not conformally equivalent to Einstein
metrics (Subsection 3.1). A particularly interesting subclass is the one with a
constant conformal scalar curvature : the constant has to be zero but the
scalar curvature is not constant. Some details on these (non-compact) metrics
are given in Subsection 4.1. Another subclass is the one with a constant scalar
curvature first
analysed by Madsen \cite{Madsen-a} : we prove that the constant has to be zero,
leading to the Einstein-Weyl structure given in ref. \cite{Madsen-a}. \newline
Second, in Subsection 3.2 we exhibit the most general conformally Einstein
metric in the family of biaxial  K\"ahler Bianchi IX metrics : the distance is
a three-parameter dependent one. As explained in Subsection 4.2, due to their
U(2) isometries, they fall into the class  proposed by Carter \cite{Carter} and
are related to extremal Calabi metrics \cite{Calabi-extremal} by the explicit
change of coordinates given by Chave and Valent \cite{CValent-extremal}.

\section{Bianchi IX metrics and Einstein-Weyl structures}
Bianchi  metrics \cite{{DS95},{Tod95}} are real four dimensional metrics with a
3-dimensional isometry group, transitive on 3-surfaces. A complete
classification has been given in 1897 by Bianchi and their expression is
usually written as :
\beqa\label{bianchi}
g & = & (abc)^2 dt^2 + a^2\si_1^2 + b^2\si_2^2 + c^2\si_3^2 \nnb \\
& = & \om_1\om_2\om_3 dt^2 + \frac{\om_2\om_3}{\om_1}\si_1^2 +
\frac{\om_1\om_3}{\om_2}\si_2^2 + \frac{\om_1\om_2}{\om_3}\si_3^2 \ ,\eeqa
where a, b, c ( or $\om_1, \om_2, \om_3$) are functions of t, and the $\si_i$
are invariant one-forms satisfying :
\beqa\label{oneforms}
d\si_1 & = & n_1\si_2\wedge\si_3 \nnb \\
d\si_2 & = & n_2\si_3\wedge\si_1 - \tilde{a}\si_1\wedge\si_2 \nnb \\
d\si_3 & = & n_3\si_1\wedge\si_2  -  \tilde{a}\si_1\wedge\si_3 \ .
\eeqa
The $n_a$'s and $\tilde{a}$ are constant parameters the possible values of
which may be found for example in refs. \cite{DS95,Tod95}).
We shall restrict our analysis to the so-called  diagonal Bianchi IX
family\footnote{\  One may always diagonalise the metric at a particular $t_0$
; then we have to assume that the evolution equations are such that the
non-diagonal components remain zero.}, characterised by
$$n_a = 1,\ \ \ a=1,2,3\ \ ;\ \ \ \tilde{a} = 0\ .$$ Due to its SU(2) isometry
group, it has been frequently considered.

A Weyl space \cite{PS93} is  a conformal manifold with a torsion-free
connection D and a one-form $\ga$ such that for each representative metric g in
a conformal class [g],
\beq\label{a1}
D_i g_{\,j\,k} = \ga_i g_{\,j\,k}\ .
\eeq
A different choice of representative metric : $g\ \ra\ \tilde{g} = e^f g$ is
accompanied by a change in $\ga\ : \ga\ \ra\ \tilde{\ga} = \ga + df\ .$
Conversely, if the one-form $\ga$ is exact, the metric g is conformally
equivalent to a Riemannian metric $\tilde{g}$ : $ D_i\tilde{ g}_{\,j\,k} = 0.$

\noindent The Ricci tensor associated to the Weyl connection D is defined by :
\beq\label{a2}
[D_i,\,D_j\,]v^k = {\cal R}^{(D)k}_{\ \ \ \ l,ij}\ v^l\ \ ,\ \ {\cal
R}^{(D)}_{ij} = {\cal R}^{(D)k}_{\ \ \ \ i,kj}\ .
\eeq
${\cal R}^{(D)}_{ij}$ is related to  $R^{(\nabla)}_{ij}$, the Ricci tensor
associated to the Levi-Civita connection \cite{PS93}:
\beq\label{a3}
{\cal R}^{(D)}_{ij} =  R^{(\nabla)}_{ij} +  \frac{3}{2}\nabla_j\ga_i -
\frac{1}{2}\nabla_i\ga_j + \frac{1}{2}\ga_i\ga_j + \frac{1}{2} g_{\,i\,j}
[\nabla_k\ga^k - \ga_k\ga^k ]\ .
\eeq
As a consequence, the Ricci tensor is no longer symmetric as :
\beq\label{a4}
{\cal R}^{(D)}_{[ij]}  = - 2\nabla_{[i}\ga_{j]}\ .
\eeq
In this work, we shall be interested in Einstein-Weyl spaces defined by :
\beqa\label{a5}
{\cal R}^{(D)}_{(ij)}  & = & \La ' g_{\,i\,j}\ \ \Leftrightarrow \nnb \\
 R^{(\nabla)}_{ij} +  \nabla_{(i}\ga_{j)} + \frac{1}{2}\ga_i\ga_j & = & \La\
g_{\,i\,j}\ \ ,\ \ \La  = \La' - \frac{1}{2}[\nabla_k\ga^k - \ga_k\ga^k] \ .
\eeqa
Notice that for an exact Einstein-Weyl structure, the representative metric is
conformally Einstein. Notice also that the conformal scalar curvature is
related to the scalar curvature through:
\beq\label{a6}
S^{(D)} = g^{\,i\,j}{\cal R}^{(D)}_{ij}  = 4\La' = S^{(\nabla)} +
3[\nabla_i\ga^i - \frac{1}{2}\ga_i\ga^i]\ .
\eeq
Diagonal Bianchi IX metrics being invariant under SU(2), it is natural to look
for one-forms $\ga$ covariant under SU(2) :
\beq\label{a7}
\ga = P(t)dt + A_1(t)\si_1 + A_2(t)\si_2 + A_3(t)\si_3\ .
\eeq
It is then proven in \cite{{Madsen-a},{Madsen-97}} that the metric is either
conformally Einstein :
\beq\label{triaxial}
A_a(t) = 0\ ,\ \ \ a=1,2,3\ ,\ \ \ga = P(t)dt\ .
\eeq
or biaxial if at least one of the $A_a(t)$ does not vanish, say $A_3$:

\beq\label{a71} \om_1 = \om_2 \ \ ,\ \ A_1 = A_2 = 0\ .
\eeq
In the following, when the three $A_a$'s vanish, we shall also restrict
ourselves to this hypothesis of U(2) isometry.

It is convenient to introduce the auxiliary functions $\al_a$ through
$$ \om'_a = \om_b\om_c + \al_a\om_a \ \ ,\ \ \ (a,b,c)\
 =\ cir.\ perm.\ (1,2,3) $$
where a prime means derivation with respect to t. Equation (\ref{a71}) implies
 $$\al_1\ =\ \al_2\ .$$
Moreover, as we shall be interested only in K\"ahler metrics, we may use a
theorem of Dancer and Strachan \cite{DS94Camb}
to obtain $$\al_3 = 0\ .$$

To sum up, when we look for an Einstein-Weyl structure with a K\"ahler diagonal
Bianchi IX metric, we have 5 unknown functions $\om_1(t) = \om_2(t)\ ,
\om_3(t), P(t), A_3(t)$ and $\al_1(t) = \al_2(t) := \al(t)$  that define the
metric, the Weyl one-form $\ga$, the K\"ahler form $\Om$ , the scalar
curvatures and the Einstein function (cosmological term) $\La$:
\beqa\label{a8}
g & = & \om_1\om_2\om_3\, dt^2 + \frac{\om_2\om_3}{\om_1}\si_1^2 +
\frac{\om_1\om_3}{\om_2}\si_2^2 + \frac{\om_1\om_2}{\om_3}\si_3^2 \nnb\\
\ga & = & Pdt + A_3\si_3 \nnb\\
\Om & = & d\om_3\wedge\si_3 + \om_3\si_1\wedge\si_2 \ \ ,\ d\Om = 0\nnb\\
S^{(D)} & = & 4\La +\frac{2}{\om_1\om_2\om_3}[P' -P^2 -\om_3^2(A_3)^2] \\
S^{(\nabla)} & = & \frac{-2\al'}{\om_1\om_2\om_3} \nnb\\
\La & = & \frac{1}{2\om_3^2}(P -2\al)\nnb
\eeqa
with 5 constraints through first order differential
equations first obtained by Madsen \cite{{Madsen-a},{Madsen-97}}(the last three
come from the diagonal matrix elements of equation (\ref{a5}) \cite{Madsen-a}:
\beqa\label{a9}
\al & = & \frac{\om'_1}{\om_1} - \om_3\ ,\ \ \ \om'_3 = (\om_1)^2\ , \nnb\\
\al' & = & 2\frac{(\om_1)^2}{\om_3}\al + P(\al +\om_3 -
\frac{(\om_1)^2}{\om_3}) + \frac{1}{2} (\om_3)^2(A_3)^2\ ,\nnb\\
P' & = & -\frac{1}{2}P^2 + 2P(\al +\om_3) +  \frac{1}{2}(\om_3)^2 (A_3)^2 \
,\nnb\\
A'_3 & = & A_3[2(\al +\om_3) - \frac{(\om_1)^2}{\om_3} - P]\ .
\eeqa
As a consequence, it is expected that everything should be fixed, up to some
integration constants.

For further use, the special cases of a (anti)self-dual Weyl connection will be
interesting.

\noindent As :
$$d\ga = A'_3dt\wedge\si_3 + A_3\si_1\wedge\si_2\ =\
\frac{1}{abc}[\frac{A'_3}{c}e^0\wedge e^3 + cA_3 e^1\wedge e^2]\ ,$$
where $e^{i}\ ,\ i=0,1,2,3$ are the vierbeins corresponding to the metric
(\ref{bianchi}), these connections are defined by :
\beq\label{a10}
A'_3 = \pm \frac{\om_1\om_2}{\om_3}A_3\ .
\eeq

\section{Solution of the constraints.}

\noindent As proposed by Pedersen and Poon \cite{PD90}, due to the positivity
of $\om'_3$, one may change the variable t to r  : $\om_3 = r^2/4$. We rewrite
the system (\ref{a8},\ref{a9}) with the new quantities :
\beq\label{b1}
\om_3 = r^2/4\ \ \ \ ,\ \ \ \ \ \ \left(\frac{\om_1}{\om_3}\right)^2 = W(r)\ ,\
W > 0\ \ .
\eeq
Then  (a dot means a differentiation with respect to r) :
\beqa\label{b2}
g & = & \frac{1}{W}dr^2 + \frac{1}{4}r^2(\si_1^2 + \si_2^2 + W\si_3^2)\ ,\ \ \
\ \ \ \ga = \frac{8 P}{r^3 W}dr + A_3\si_3\ , \nnb\\
\Om & = & \frac{1}{4}(dr^2\wedge\si_3 + r^2\si_1\wedge\si_2)
\ ,\ \ \ \ \ \al = \frac{1}{16}r^3\dot{W} -\frac{1}{4}r^2(1-W)\ ,\nnb\\
S^{(D)} & = & -\frac{4(A_3)^2}{r^2 W}- \frac{192P^2}{r^6 W} +\frac{96P}{r^4}
+\frac{16P\dot{W}}{r^3 W} - \frac{4\dot{W}}{r} +\frac{16(1-W)}{r^2}\ ,\nnb \\
S^{(\nabla)} & = & -\frac{16\dot{\al}}{r^3}\ ,\ \ \ \ \ \La = \frac{8}{r^4}(P
-2\al)\ ,
\eeqa
and the constraints, also given in \cite{{Madsen-a},{Madsen-97}} write :
\beqa\label{b3}
\dot{\al} & =  & \frac{r(A_3)^2}{4W} + \frac{\dot{W}P}{2W} +r(
\frac{r\dot{W}}{4} +W -1)\ ,\nnb \\
\dot{P} & = & -\frac{4P^2}{r^3W} +  P(\frac{\dot{W}}{W} + \frac{4}{r}) +
\frac{r(A_3)^2}{4W}\ ,\\
\dot{A_3} & = & A_3\left[\frac{\dot{W}}{W} + \frac{2}{r} - \frac{8 P}{r^3
W}\right]\ .\nnb
\eeqa
Notice also that with the new variable r, the (anti)self-duality of the Weyl
connection (\ref{a10}) writes :
\beq\label{a11}
\dot{A_3} = \pm \frac{2}{r}A_3 \Leftrightarrow \ \ A_3 = \la r^{\pm 2}\ ,
\eeq

Coming now to the system (\ref{b3}), we distinguish 3 cases, corresponding
to $A_3$ or $P$ being zero or not.
\subsection{Non exact Weyl form : $A_3 \neq 0$.}

Let us define new functions by :
$$R(r) = \frac{W(r)}{A_3(r)}\ \ ,\ \ \ \ T(r) = \frac{P(r)}{r^2 A_3(r)}\ .$$
The second and third equations (\ref{b3}) give :
\beq\label{e3}
T = R/4 + r\dot{R}/8\ ,
\eeq
\beq\label{e3b}
 2r^2R\ddot{R} - (r\dot{R})^2 + 2rR\dot{R} - 4R^2 - 4 = 0\ ,
\eeq
Thanks to (\ref{e3}), the Weyl form becomes :
\beq\label{e3c}
\ga = \frac{8T}{rR}dr + A_3 \si_3 = d\log \mid r^2R\mid +
A_3 \si_3\ .
\eeq
Notice the invariance of the equations in the simultaneous change of R, T and
$A_3$ in their opposite.
The non-linear, but homogeneous, differential equation (\ref{e3b}) may be
integrated  once to give ($\be$ being an arbitrary constant)$$r\dot{R} =
2\epsilon\sqrt{R^2 + 2\be R - 1} \equiv 2\epsilon\sqrt{(R-a_1)(R+a_2)}\ \ , \ \
\epsilon = \pm 1\ .$$
A further integration gives  ($d_{\epsilon}$ being other strictly positive
arbitrary constants)
$$\log(d_{\epsilon} r) =  \epsilon\log[ \sqrt{R_{\epsilon}-a_1}+
\sqrt{R_{\epsilon}+a_2}]\ .$$
The function R(r) is then obtained :
\beq\label{b31}
 R_{\epsilon}(r) = -\be + \frac{(d_{\epsilon}  r)^{2\epsilon}}{8} +
2\frac{(1+\be^2)}{(d_{\epsilon} r)^{2\epsilon}}\ ,
\eeq
but there is a positivity constraint that writes
$$  (d_{\epsilon}r)^{2\epsilon} \ge 4\sqrt{1+\be^2}\ .$$
So, $R_{+}$ is defined for $(d_{+}r)^{2} \ge 4\sqrt{1+\be^2}$ and $R_{-}$ for
$(d_{-}r)^{2} \le \frac{1}{4\sqrt{1+\be^2}}\ .$ A continuous solution of
(\ref{e3b}) is  obtained iff. $d_{-} = \frac{d_{+}}{4\sqrt{1+\be^2}}\ ,$
and writes, using the variable $\tau = d^2 r^2 /4\ :$
\beq\label{31b}
 R(r) = -\be + \frac{\tau}{2} + \frac{(1+\be^2)}{2\tau}\ =
\frac{1}{2\tau}[(\tau - \be)^2 +1]\ .
\eeq
Equation (\ref{e3}) then gives
\beq\label{31c}
T(r) = \frac{1}{4}[\tau - \be]\ .
\eeq

\noindent Differentiating $\al(r) $ given in equ.(\ref{b2}) with respect to r,
when compared to the first equ.(\ref{b3}), gives a second order linear
differential equation on  $G(\tau) \equiv A_3(r)\ :$
\beq\label{b32}
\frac{\tau^2}{2}\frac{d^2G}{d\tau^2}\left[(1+\be^2) - 2\be\tau +\tau^2\right] +
\tau\frac{dG}{d\tau}\left[-\be\tau +\tau^2\right] - G\left[(1+\be^2) - \be\tau
+\tau^2\right] = - 2\tau\ .
\eeq
Using $G = 2/\tau + \la_1[\tau/4 - (1+\be^2)/(4\tau)]$ as one parameter family
of solutions ( easily found by inspection of the recursion relation associated
to a Laurent series expansion of G around 0), one finds the general solution of
(\ref{b32}) :
\beq\label{b34}
G = \frac{2}{\tau} + \frac{\la_1}{4\tau}[\tau^2 - (1+\be^2)] +
\frac{\la_2}{8\tau}\left[\tau + \be + [\tau^2 -(1+\be^2)] \arctan{(\tau -
\be)}\right]\ .
\eeq
Furthermore, thanks to the previous remark, a $\pm 1$ factor may be introduced,
simultaneously, in R(r), T(r) and $A_3(r)\ .$

As a consequence, we have the following Theorem :

\vspace{0.3cm}

\noindent {\bf Theorem 1 :} {\sl Any diagonal K\"ahler Bianchi IX metric $g$
and non-exact Weyl one-form $\ga$ such that  the structure ($g\ ,\ \ga$) is
that of an Einstein-Weyl space, depend on 4 parameters ($\la_1,\ \la_2,\ \be$
and $d$) and are given by} :
\newline i) {\sl the metric} $g$
\beqa\label{35a}
g & = & \frac{1}{d^2}\left[\frac{d\tau^2}{\tau V(\tau)} +\frac{\tau}{4}(\si_1^2
+ \si_2^2) +
\frac{1}{4}\tau V(\tau)\si_3^2\right]\ \ ,\ \ \ \ \
\tau V(\tau) = \frac{1}{2}[1+(\tau - \be)^2]G(\tau)\ \ ,\nnb \\
\tau V(\tau) & = & \left[\tau - 2\be + \frac{1+\be^2}{\tau}\right]
+\frac{\la_1}{8}\left[\tau^3 -  2\be\tau^2  + 2\be(1+\be^2) -
\frac{(1+\be^2)^2}{\tau}\right] +\\
& + &  \frac{\la_2}{16}\left[\tau^2 - \be\tau  + (1-\be^2)
+\frac{\be(1+\be^2)}{\tau} +
\frac{[(\tau- \be)^2 +1][\tau^2 - (1+\be^2)]}{\tau}\arctan{(\tau -
\be)}\right]\ ,\nnb
\eeqa
where $\tau = \frac{d^2r^2}{4}\ ,\ \ V(\tau) \equiv W(r)\,,$
\newline ii)  {\sl the one-form} $\ga$
\beqa\label{35b}
\ga & = & d\log[1 + (\tau - \be)^2] \pm G(\tau)\si_3\ \ ,\nnb \\
G(\tau) & = &  \frac{2}{\tau} + \frac{\la_1}{4\tau}[\tau^2 -(1+\be^2)] +
\frac{\la_2}{8\tau}\left[\tau + \be + [\tau^2 - (1+\be^2)]\arctan{(\tau -
\be)}\right]\ ;
\eeqa
\newline iii) {\sl the geometrical quantities result from  equation}
(\ref{b2}),
\beqa\label{b35c}
\frac{1}{d^2}S^{(D)} & = & -\frac{\la_2}{2[1+(\tau -\be)^2]}\nnb \\
\frac{1}{d^2}S^{(\nabla)} & = & - \frac{3\la_1}{2}[\tau - \be] -
\frac{3\la_2}{4}\left[[\tau - \be]\arctan{(\tau-\be)} + \frac{(\tau - \be)^2 +
2/3}{(\tau - \be)^2 + 1} \right]\nnb \\
d^2\al & = & -\be + \frac{\la_1}{8}[\be(1+\be^2) - 3\be\tau^2 + 2\tau^3] + \\
 & + & \frac{\la_2}{16}\left[ (\tau - \be)(2\tau +\be) + [\be(1+\be^2) -
3\be\tau^2 + 2\tau^3]\arctan{(\tau - \be)}\right]\ , \nnb \\
\frac{1}{d^2}\La & = & \frac{1}{\tau}
 -\frac{\la_1}{8\tau}[(\tau - \be)^2 + 1]  - \frac{\la_2}{16\tau}\left[\tau -
\be +[ 1 + (\tau - \be)^2]
\arctan{(\tau - \be)}\right]\ .\nnb
\eeqa
$$  $$
A few comments and corollaries are in order :
\begin{itemize}
\item The expression of the conformal scalar curvature
exhibits a well-defined sign, depending only on the parameter $\la_2\,.$ For
4-dimensional compact Einstein-Weyl spaces, this results from the work of
Pedersen and Swann
\cite{PDSRheine}. In particular, for the compact case, if  $S^{(D)}$ vanishes,
the space is conformally Einstein.

\item In the basis of self-dual two-forms, the self-dual part of the Weyl
curvature tensor writes:
$$W^{+} = \frac{S^{(\nabla)}}{12}\left[\begin{array}{ccc}
-1 & 0 & 0 \\
0 & -1 & 0 \\
0 & 0 & +2 \\
\end{array}\right]$$
\item The condition $S^{(D)} = $ constant implies
$S^{(D)} = 0$ which is equivalent to $\la_2=  0\ .$
This leads to the following corollary :
\end{itemize}
{\bf Proposition 1} : {\sl  Given a diagonal K\"ahler Bianchi IX metric $g$ and
a Weyl one-form $\ga$ such that  the structure ($g\ ,\ \ga$) is that of an
Einstein-Weyl space, the two following conditions are equivalent :
\newline i) the space has a constant conformal scalar curvature,}
\newline ii) {\sl the metric, depending on 3 parameters submitted to some
positivity requirements} \footnote{\ To agree with commonly used notations, we
make the following changes of parameters :
$$\ka/6 = - \frac{\la_1 \be d^2}{16}\ ,\ \ 8\la = -\frac{\be}{d^2}[8 -
\la_1(1+\be^2)]\ ,\ \ 16\mu = \frac{2(1+\be^2)}{d^4}[8 - \la_1(1+\be^2)]\ .$$
Then the quantity $[\mu - \la^2 -\frac{\ka\mu^2}{3\la}]/\la^2$ equals the
positive constant $1/\be^2\,;$ this, as we shall see in Subsect. 4.1 needs
$\ka\la\ <\ 3/4\ .$ Moreover, W should be a positive function.}, {\sl is}
\beqa\label{e7}
g & = & \frac{1}{W}dr^2 +\frac{1}{4}r^2(\si_1^2 + \si_2^2 + W\si_3^2)  \\
W(r) & = & 1 + \frac{\ka r^2}{6} +\frac{\ka(3\la  - \ka\mu)r^4}{144\la^2} +
\frac{8\la}{r^2} +   \frac{16\mu}{r^4} = \left(2\mu + \la r^2
+\frac{3\la-\ka\mu}{24\la}r^4\right)\left(\frac{8}{r^4} +
\frac{\ka}{6\la}\right)\ , \nnb
\eeqa
{\sl the conformal scalar curvature vanishes and the one-form and the scalar
curvature are respectively :}
\beqa\label{e8}
\ga & = & d\log[2\mu + \la r^2 +\frac{3\la-\ka\mu}{24\la}r^4]\ \pm\
\si_3\sqrt{[\mu - \la^2 -
 \frac{\ka\mu^2}{3\la}]}\left[\frac{8}{r^2} + \frac{\ka r^2}{6\la}\right]\nnb
\\
S^{(\nabla)} & = & -4\ka[1 + \frac{3\la - \ka\mu}{12\la^2}r^2]\ .
\eeqa
\begin{itemize}
\item The condition of constant scalar curvature enforces $\la_1 = \la_2 =  0\
,i.e.\ S^{(\nabla)} = 0\,.$ This leads to a special case ($\ka = 0$) of the
previous proposition which appears to be the scalar-flat metric of Madsen with
an anti-self dual Weyl connection  (see
 equ.(\ref{a11})) \cite{{Madsen-a},{Madsen-97}}. Then we have the following
corollary :
\end{itemize}
{\bf Proposition 2} : {\sl  Given a diagonal K\"ahler Bianchi IX metric $g$ and
a Weyl one-form $\ga$ such that  the structure ($g\ ,\ \ga$) is that of an
Einstein-Weyl space, the two following conditions are equivalent :
\newline i) the space has a constant scalar curvature,}
\newline ii) {\sl the metric, Weyl form and geometric parameters, depending on
2 parameters submitted to} $\mu \ge \la^2$, {\sl are} :
\beqa\label{cc3}
g & = & \frac{1}{W}dr^2 +\frac{1}{4}r^2(\si_1^2 + \si_2^2 + W\si_3^2)\ ,\ \ \ W
= 1 + \frac{8\la}{r^2} + \frac{16\mu}{r^4} = (1+\frac{4\la}{r^2})^2 +
\frac{16(\mu - \la^2)}{r^4}\ >\ 0\ ,\nnb \\
\ga & = & d\log[r^4W] \pm \frac{8\sqrt{\mu - \la^2}}{r^2}\si_3 \ ,\ \ \ \ W^{+}
= S^{(\nabla)} = S^{(D)} = 0\ .
\eeqa
This structure has been recently related to Ricci-flat metrics with torsion
\cite{Todd96}.
\begin{itemize}
\item The special case $\la_2 = 0,\ \ \la_1 = 8/(1+\be^2)$ leads to :
$$W = 1 - \frac{\be d^2}{2(1+\be^2)}r^2 + \frac{d^4}{16(1+\be^2)}r^4\ \ ,\ A_3
= \pm\frac{d^2}{2(1+\be^2)}r^2\ ,$$
which is an Einstein-Weyl structure with a self-dual Weyl connection (see
equ.(\ref{a11})) :
\beqa\label{cc4}
W & = & 1 + \frac{\ka}{6}r^2 + \frac{\nu}{48}r^4 = (1 + \frac{\ka r^2}{12})^2
+\frac{(3\nu - \ka^2)r^4}{144}\ >\ 0\ ,\ \ \nu >  \ka^2 /3\ ,\nnb \\
\ga & = & d\log[W] \pm\sqrt{3\nu - \ka^2}\left(\frac{r^2}{6}\right)\si_3\ ,\ \
\ S^{(D)} = 0\ ,\ \ \ S^{(\nabla)} = -(4\ka +\nu r^2)\ .
\eeqa

\item By inspection of expression (\ref{35b}) and use of equation (\ref{a11}),
we have the following corollary:
\end{itemize}
{\bf Proposition 3} : {\sl  Given a diagonal K\"ahler Bianchi IX metric $g$ and
a non exact Weyl one-form $\ga$ such that the structure ($g\ ,\ \ga$) is that
of an Einstein-Weyl space, the only (anti)self dual Weyl connections are those
of equations (\ref{cc3}) and (\ref{cc4}).}
\vspace{.5cm}

Some properties of the conformally scalar flat metric (\ref{e7}) will be
offered in  Subsection 4.1.

\subsection{Conformally Einstein metrics : $A_3 = 0\ ,P\neq 0\,.$}

As said before, we consider this case as a limit of the general class, and we
still ask $\om_1 = \om_2$ (biaxial or U(2) metrics), which is definitely a
restriction (see for
 example see some attempts in \cite {PD90}).

\noindent We add to the Einstein-Weyl conditions (\ref{b2},\ref{b3}), the
exactedness of the Weyl one-form $\ga$,
$$\ga = d\tilde{\ga}\ \ \Leftrightarrow \ A_3(r)=0\ .$$
A first constraint writes :
\beq\label{d1}
\frac{\dot{P}}{P} = \frac{\dot{W}}{W} +\frac{4}{r} - \frac{4P}{r^3 W}\ .
\eeq
The function $P(r)$ being different from zero (to have a non vanishing Weyl
form), we define an auxiliary function $$R(r) = \frac{r^4 W(r)}{P(r)}\ ;$$
it satisfies :
\beq\label{d11}
\dot{R} = 4r
\eeq
which gives ($\de$ being an arbitrary constant) :
\beq\label{d2}
P(r) = W(r)\frac{r^4}{2(r^2 + 4\de)}\ .
\eeq
Notice that, thanks to (\ref{d11}), the Weyl form writes :
$$\ga = \frac{8r}{R} = 2d\log{R}$$
The constraint (\ref{b3}) on the function $\al(r)$ of (\ref{b2}) then gives
$$ \ddot{W}\frac{r^2}{16} - \dot{W}\left[\frac{r^3}{4(r^2 + 4\de)}
-\frac{3r}{16}\right] -\frac{1}{2}(W-1) =0\ ,$$
which writes, using the variable $\tau = r^2$ and the function $V(\tau) = W(r)
- 1$
\beq\label{d3}
\tau ^2(\tau + 4\de)\ddot{V} +8\de\tau\dot{V} -2(\tau + 4\de)V = 0\ .
\eeq
This linear equation integrates to
\beq\label{d4}
V = a_1\tau(1 + \frac{\tau}{8\de}) + b_1\frac{(\tau +2\de)}{\tau^2}\ ,
\eeq
and this gives
\beq\label{d5}
W(r)  = 1 +  a_1r^2 +\frac{a_1}{8\de}r^4 + \frac{b_1}{r^2}
 + \frac{2\de b_1}{r^4}\ .
\eeq
With the following change of parameters : $a_1 = \ka/6\,,b_1 = 8\la\,$ and $\de
= \mu/\la\ ,$ we have the
following theorem :
$$  $$
{\bf Theorem 2} : {\sl Given a diagonal K\"ahler Bianchi IX metric $g$ and a
Weyl one-form $\ga$ such that  the structure ($g\ ,\ga$) is that of an
Einstein-Weyl space, the two following conditions are equivalent} :
\newline i){\sl The space is conformally equivalent to an Einstein space, and
the metric is biaxial,}
\newline ii) {\sl The metric depends on 3 arbitrary parameters }:
\beq\label{d6}
g = \frac{1}{W}dr^2 +\frac{1}{4}r^2(\si_1^2 + \si_2^2 + W\si_3^2)\ \ ,\ \ \
W(r)  = 1 + \frac{\ka r^2}{6} +\frac{\ka\la r^4}{48\mu} +  \frac{8\la}{r^2} +
\frac{16\mu}{r^4}
\eeq
{\sl and the one-form is}
\beq\label{d7}
\ga = d\log[4\mu + \la r^2]^2\ .
\eeq

\noindent Moreover, with (\ref{b2}),
\beqa\label{b36}
S^{(D)}  = \frac{192[\la\mu-\la^3 -\frac{\ka\mu^2}{3}]}{[4\mu + \la r^2]^2}\ \
\ & , &  S^{(\nabla)} = - 4\ka[1 + \frac{\la r^2}{4\mu}]\ , \\
\al = \la + \frac{\ka r^4}{16} + \frac{\la\ka r^6}{96\mu}\ \ \ & , & r^2\La  =
\frac{16}{4\mu + \la r^2}[\la^2  + \frac{\la-\ka\mu}{4}r^2 -
\frac{\ka\la}{16}r^4  - \frac{\ka\la^2}{192\mu}r^6]\nnb \ .
\eeqa
A few comments are in order :
\begin{itemize}
\item The request of a constant scalar curvature leads either to $\la = 0$,
which is an Einstein metric and will be discussed in the
next subsection, or $\ka = 0$ which is scalar flat (see the
 next item).

\item The special case $a_1 = 0 \Rightarrow \ka = 0$, where the scalar
curvature vanishes, gives a metric conformally equivalent to the Lebrun metric
\cite{Lebrun} which generalises Eguchi-Hanson ($\ka = \la =0$).

\item The special case $b_1 = 0 \Rightarrow \la = \mu =0$ with a fixed ratio
$\de$ leads to :
\beqa\label{b37}
W(r) = 1 + (\ka/6)r^2 + (\ka/48\de)r^4\ \ \ & ,& \ga = d\log[4\de + r^2]^2\
,\nnb \\
S^{(D)} = \frac{-64\de(\de\ka -3)}{[4\de + r^2]^2}\ \ \ & ,&  S^{(\nabla)} =  -
4\ka[1 + \frac{r^2}{4\de}]\ .
\eeqa
With $\nu = \frac{\ka}{\de}$, we recover the metric
(\ref{cc4}), but the Weyl one-form and the conformal scalar curvature are
different.

\item The special case $b_1 = 0,\ \de\ \infty $ such that $b_1 \de = \mu
\Rightarrow \la = 0$, is the
aforementioned Einstein metric.

\item The special case $a_1 = \de =0 $ with a  fixed ratio $a_1/\de = \nu/6
\Rightarrow \ka = \mu =0$ with a fixed ratio $\ka/\mu = \nu/\la$ leads to :
\beqa\label{b38}
W(r) = 1 + (8\la)/r^2 +(\nu/48)r^4 \ \ \ &, &\ga = d\log[r^2]^2\ ,\nnb \\
S^{(D)} = \frac{-192\la}{r^4}\ \ \ &, & S^{(\nabla)} = - \nu r^2\ .
\eeqa

\item The special case
\beq\label{b39}
b_1 = 8\de[1 - 2a_1\de]\ \Rightarrow\ [\mu/\la-\la -\frac{\ka\mu^2}{3\la^2}] =
0\ \Rightarrow \ S^{(D)} = 0\ ,
\eeq
is common to this subsection and the previous one : under this hypothesis, the
two quantities $\frac{\ka(3\la-\ka\mu)}{144\la^2}$ in
 equ.(\ref{e7})(Proposition 1) and $\frac{\ka\la}{48\mu}$ in
equ.(\ref{d6})(Theorem 2) are indeed the same.

\end{itemize}
In Subsection 4.2, we shall analyse the Einstein metrics conformally equivalent
to (\ref{d6}) and recognise previously known ones.

\subsection{Einstein metrics : $A_3 = P = 0\,.$}
The constraint (\ref{b3}) on the function $\al(r)$ of (\ref{b2}) gives the
homogeneous equation
$$ r^2\ddot{W} + 3r\dot{W} - 8(W-1) =0\ ,$$
which integrates to
\beq\label{d41}
W = 1 + a_2 r^2 + b_2/r^4\ ,
\eeq
that was obtained by Gibbons and Pope \cite{Gibbons79} and Pedersen
\cite{Pedersen85} (It interpolates between Eguchi-Hanson ($a_2 = 0$) and the
pseudo Fubini-Study metric ($b_2 = 0$).) It is the only K\"ahler Einstein
biaxial Bianchi IX metric \cite{DS94Camb}. The scalar curvatures $S^{(D)}$ and
$S^{(\nabla)}$ - as well as the cosmological term $\La$ - are both equal to the
constant $-6a_2\,.$

\section{More on  our Einstein-Weyl metrics}
\subsection{Conformally scalar flat metrics}

In the present work, we comment only the conformally scalar flat special case
($\la_2 = 0$ in equation (\ref{35a}), i.e. the result of Proposition 1). A
first remark is that such spaces cannot be compact, because according to a
result of Pedersen and Swann \cite{PDSRheine}, they should then be conformally
Einstein.

\noindent When rewritten as a function of $\tau = r^2\,,$ the distance
\footnote{\ The expression (\ref{e77}) of the distance appears to be the same
as the one obtained in \cite{CValent-extremal} for the extremal metrics of
Calabi \cite{Calabi-extremal}, and compared to equation 8 of
\cite{CValent-extremal},  satisfies  the relation
$$c_1 c_3^2 + c_2 c_3 + c_2^2 c_4 = 0\ .$$} and Weyl form are :
\beqa\label{e77}
g & = & \frac{1}{4}\left[\frac{1}{\rho(\tau)}d\tau^2 + \tau(\si_1^2 + \si_2^2)
+ \rho(\tau)\si_3^2\right]\ ,\ \ with \ \ \ \rho(\tau) \equiv \tau W(r)\nnb \\
\rho(\tau)  & = &  \frac{16\mu}{\tau} + 8\la + \tau + \frac{\ka}{6}\tau^2
+\frac{\ka(3\la -\ka\mu)}{144\la^2}\tau^3 = \left(2\mu + \la\tau
+\frac{3\la-\ka\mu}{24\la}\tau^2\right)\left(\frac{8}{\tau} +
\frac{\ka}{6\la}\tau\right)\ ;\nnb \\
\ga & = & d\log[2\mu + \la\tau +\frac{3\la-\ka\mu}{24\la}\tau^2]\ \pm\
\si_3\sqrt{[\mu - \la^2 -
 \frac{\ka\mu^2}{3\la}]}\left[\frac{8}{\tau} +  \frac{\ka}{6\la}\tau\right]\  .
\eeqa
Remember that the quantity $ \mu - \la^2 -
 \frac{\ka\mu^2}{3\la}$ has to be positive. One easily  finds that this
requires $$\ka\la \le3/4\ .$$
One gets :
$$\mu \le \frac{3\la}{2\ka}[\sqrt{1-\frac{4\ka\la}{3}} + 1]\
\le 0\ \ \ \ {\rm or}\ \ \ \  \mu \ge
-\frac{3\la}{2\ka}[\sqrt{1-\frac{4\ka\la}{3}} - 1]\ \ge 0\ ,$$
if $\ka\la \le 0$, and, if $0 \le \ka\la \le 3/4$ :
$$ 0 \le \frac{3\la}{2\ka}[1 - \sqrt{1-\frac{4\ka\la}{3}}]\ \ \le \mu\
\ \le \frac{3\la}{2\ka}[1 + \sqrt{1-\frac{4\ka\la}{3}}]\ .$$
Moreover, the first parenthesis of $\rho(\tau)$ given in equ.(\ref{e77}) cannot
vanish and has the same sign as $\mu\ .$ Let us now discuss the positivity of
the metric. The second parenthesis in
$\rho$ may vanish, for $\tau = \tau_{0} = 4\sqrt{\frac{-3\la}
{\ka}}$, iff. $\ka\la \le 0\ .$ Finally, three situations arise :
\begin{itemize}
\item a) $0 \le \ka\la \le 3/4\ :$ the parameter $\mu$ is positive (restricted
as mentioned above) and the function $\rho(\tau)$ too. The variable $\tau$
belongs to the interval $]0,\ \infty[\ ;$
\item b) $\ka\la \le 0$ and $ \mu \ge
\frac{-3\la}{2\ka}[\sqrt{1-\frac{4\ka\la}{3}} - 1] \ge 0\ :$ the function
$\rho(\tau)$ - and the metric - is positive in the interval $]0,\ \tau_{0}[\ ;$
\item c) $\ka\la \le 0$ and $ \mu \le
\frac{3\la}{2\ka}[\sqrt{1-\frac{4\ka\la}{3}} + 1] \le 0\ :$ the function
$\rho(\tau)$ - and the metric - is positive in the interval $]\tau_{0},\
\infty[\ .$
\end{itemize}

Let us now discuss the kind of singularities of the metric
with respect to the terminology of Gibbons and Hawking \cite{GibHaw}.
The value $\tau = 0$ is an incurable singularity except when  $\mu = 0$ (then
$\la$ has to vanish  and one is led to the particular self-dual case
(\ref{cc4}) where, the function $W$ being positive definite, $\tau$ belongs to
$]0,\ \infty[$ and the other end, $+\infty$, is singular).
The value $\tau = +\infty$ is an incurable singularity except when $\la =
\ka\mu/3$ (then $\ka$ has to vanish and one is led to the particular
antiself-dual case  (\ref{cc3}) where, the function $W$ being positive
definite, $\tau$ belongs to $]0,\ \infty[$ and the other end, $0$, is
singular).

\noindent With respect to the  singularity at $\tau_{0}$, we now prove that,
only in subcase c), it may be an  apparent bolt-like singularity . $\rho(\tau)$
being positive, the expansion
$$\rho(\tau) = (\tau - \tau_{0})\rho'(\tau_{0}) +...$$
allows the change of variable :
$$\tau - \tau_{0} = x^2 \rho'(\tau_{0})\ ,$$
and the metric behaves like
$$ (dx)^2 + x^2[\frac{\rho'(\tau_{0})}{2}]^2\si_3^2 + \frac{1}{4}[\tau_{0} +x^2
\rho'(\tau_{0})](\si_1^2
+ \si_2^2) +..$$
exhibiting a bolt \cite{GibHaw} of order k iff. :
$$\frac{\rho'(\tau_{0})}{2} = k\ \ ,\ \ \ k=1,2,...\ \ ;$$
indeed, in such a case, the singularity at $\tau_{0}$ is an apparent
polar-coordinate like singularity, which disappears
when the range in the angle $\psi$ ($\si_3 = d\psi + \cos{\theta} d\phi,..$) is
limited to
$[0,\ \frac{4\pi}{k}]\ .$

\noindent This requires :
\beq\label{bolt1}
\mu = -\frac{k+1}{32}\tau_{0}^2 - \frac{\la}{4}\tau_{0}\ \ ,\ \ \ \ \ \ka =
-\frac{48\la}{\tau_{0}^2}\ ,
\eeq
and the positivity of $ \mu - \la^2 -\frac{\ka\mu^2}{3\la}$
writes :
\beq\label{bolt2}
\la\ \ge\ -\frac{k^2 -1}{16k}\tau_{0} \ \ \Rightarrow \ \ \mu\ \le\
-\frac{(k+1)^2}{64k}\tau_{0}^2\ <\ 0\ .
\eeq
Then, we can summarise the previous discussion according to cases a), b) and c)
defined above:
\begin{itemize}
\item a) $\tau \in\ ]0,\ \infty[$ : the metric is singular at both ends, except
for the (anti)self-dual cases mentioned above ;

\item b) $\tau \in\ ]0,\ \tau_{0}[$ : the metric is singular at both ends ;
\item c) $\tau \in\ [\tau_{0},\ \infty[$ : if the parameters are related as in
equ. (\ref{bolt1}), the singularity at $\tau_{0}$ is bolt-like, and the metric
is complete at $\tau_{0}$, but anyway singular at $\infty\,.$
\end{itemize}
As explained above, they are never \footnote{\ Except when they are in fact
conformally Einstein ($ \mu - \la^2 -\frac{\ka\mu^2}{3\la} = 0\,.$)} compact
\cite {PDSRheine}.

\subsection{Conformally Einstein metrics}

The metric g (\ref{d6}), with
$$\ga = d\log[h^2]\ ,\ \ h = 4\mu + \la r^2\ ,\ \  S^{(\nabla)}_{g} =
-\frac{\ka}{\mu}h\ ,$$
is conformally equivalent to the metric $\bar{g} = g h^{-2}$ which is Einstein,
with a scalar curvature given by :
\beq\label{f1}
 S^{(\nabla)}_{\bar{g}} = h^2  S^{(\nabla)}_{g} -
6[h\triangle h + 2\mid dh\mid^2] = 192[\la\mu - \la^3 -\frac{\ka\mu^2}{3}]\ .
\eeq
This is in agreement with a theorem of Derdzinski \cite{Der83} quoted in the
book of A. Besse (\cite{Besse},
 Theorem 11.81) as the function
$$\left(S^{(\nabla)}_{g}\right)^3 - 6[S^{(\nabla)}_{g}\triangle
S^{(\nabla)}_{g} + 2\mid dS^{(\nabla)}_{g}\mid^2] =
\frac{\ka^2}{\mu^2}S^{(\nabla)}_{\bar{g}}$$
is a constant ; moreover, this theorem proves that g is a conformally  Einstein
extremal metric of Calabi \cite{Calabi-extremal}. Then the work of Valent and
Chave \cite{CValent-extremal}, giving an explicit form for their distance and
the change of coordinates ensuring this link, is of great help.
Our U(2) symmetric metrics (\ref{d6}), when expressed as a function of $\tau =
r^2$ :
\beqa\label{f2}
g & = & \frac{\tau}{4\rho(\tau)}d\tau^2 + \frac{1}{4}\tau(\si_1^2 + \si_2^2) +
\frac{1}{4\tau}(\rho(\tau))\si_3^2)\ \ ,\\
\rho(\tau) & = & \tau^2 W(r)  = 16\mu +  8\la\tau + \tau^2 +
\frac{\ka}{6}\tau^3 +\frac{\ka\la}{48\mu}\tau^4 \ ,\nnb
\eeqa
and compared to equation 8 of \cite{CValent-extremal},  satisfies  the relation
$4c_1 c_4 = c_2 c_3$ of
 proposition 7 (equ. 15 of \cite{CValent-extremal}), so ensuring their
conformally Einstein character. With respect to $\bar{g}$, the change of
coordinate :
$$\rho = \frac{\la\tau - 4\mu}{\la\tau + 4\mu}\ \ ,\ \ \ -1 < \rho\ < +1$$
gives a Carter metric \cite{Carter}
\beqa\label{f3}
\bar{g} & = & \frac{1}{64\mu\la}\left\{ \frac{1 - \rho^2}{\Delta(\rho)}d\rho^2
+  \frac{4\Delta(\rho)}{1 - \rho^2}\si_3^2 + (1 -  \rho^2) (\si_1^2 +
\si_2^2)\right\}   \\
\Delta(\rho) & = & [1+\frac{\ka\mu}{\la} + \frac{3\la^2}{\mu}] +
\frac{8}{3}[ \frac{\ka\mu}{\la} - \frac{3\la^2}{\mu}]\rho +
2[-1+\frac{\ka\mu}{\la} + \frac{3\la^2}{\mu}]\rho^2 -
\frac{1}{3}[-3 +\frac{\ka\mu}{\la} + \frac{3\la^2}{\mu}]\rho^4\nnb\ .
\eeqa
Finally, if one require compactness and special behavior at end points
(bolt-bolt), one gets definite values for our parameters $\la, \ka$ and
$\mu\,.$ In particular,  $\bar{g}$ is the Page metric \cite{Page} iff.
$$ \frac{2\ka\mu}{3\la} = \frac{2\la^2}{\mu} = 1 -
\frac{1+\nu^2}{\nu(3+\nu^2)}\ \simeq - .2443\,$$
where the Page parameter $\nu \simeq .2817$ is the unique root of the algebraic
equation $-\nu^4 +6\nu^2 +3 = 4\nu(3+ \nu^2)$ in the range $0 < \nu <1\,.$

\section {Concluding remarks}
In this paper, we have presented a complete analysis of the  Einstein-Weyl
structures ($g\,,\ \ga$) corresponding to
 diagonal K\"ahler Bianchi IX metrics first analysed from this point of view by
Madsen. The general solution is a 4-parameter metric (\ref{35a}). In the
subclass where the conformal scalar curvature is a constant, our Proposition 1
extends previously known results
 \cite{{Madsen-a},{Madsen-97}} to  non scalar-flat cases. Unfortunately these
metrics, known to be non-compact, are always singular. We refer to a further
publication for a discussion on the link between our Einstein-Weyl structures
and geometries with torsion as initiated by Tod \cite{Todd96}.

With respect to conformally Einstein structures, we
 recover the U(2) symmetric metrics conformally equivalent to the Einstein
metrics of Carter \cite{{Carter},{CValent-extremal}}.

\bibliographystyle{plain}

\begin {thebibliography}{29}
\bibitem{Todd96} K. P. Tod, {\sl Class. Quantum Grav.} {\bf 13} (1996) 2609.

\bibitem{PS93} H. Pedersen and A. Swann, {\sl Proc. Lond. Math. Soc.} {\bf 66}
(1993) 381.

\bibitem{HullWitten85} C. M. Hull and E. Witten, {\sl Phys. Lett.} {\bf 160B}
(1985)  398 ; C. M. Hull {\sl Nucl. Phys.} {\bf B267} (1986) 266.


\bibitem{xxy} E. Bergshoef and E. Sezgin, {\sl Mod. Phys. Lett.} {\bf A1}
(1986) 191 ; \newline P. Howe and G. Papadopoulos,
{\sl Nucl. Phys.} {\bf B289} (1986) 264 ; {\sl Class. Quantum Grav.} {\bf 4}
(1987) 1749 ; {\sl Class. Quantum Grav.} {\bf 5} (1988) 1647 ; \newline Ph.
Spindel, A. Sevrin, W. Troost and A. Van Proyen, {\sl Nucl. Phys.} {\bf B308}
(1988) 662 ;
\newline  F. Delduc, S. Kalitzin and E. Sokatchev, {\sl Class. Quantum Grav.}
{\bf 7} (1990) 1567.

\bibitem{Delduc} F. Delduc and G. Valent, {\sl Class. Quantum Grav.} {\bf 10}
(1993) 1201.

\bibitem{DS95} A. S. Dancer and Ian A. B. Straham, {\sl Cohomogeneity-One
K\"ahler metrics} in ``Twistor theory", S. Huggett ed., Marcel Dekker Inc., New
York, 1995, p.9.

\bibitem{Tod95} K. P. Tod, {\sl  Cohomogeneity-One metrics with Self-Dual Weyl
tensor} in ``Twistor theory", S. Huggett ed., Marcel Dekker Inc., New York,
1995, p.171.

\bibitem{Madsen-a} A. Madsen, {\sl Compact Einstein-Weyl manifolds with large
symmetry group}, PhD. Thesis (especially section 9.2), Odense University, 1995.

\bibitem{Madsen-97} A. Madsen, {\sl Einstein-Weyl strutures in the conformal
classes of Lebrun metrics}, to be
 published in {\sl Class. Quantum Grav.}


\bibitem{Carter} B. Carter, {\sl Comm. Math. Phys.} {\bf 10} (1968) 280.

\bibitem{Calabi-extremal} E. Calabi, {\sl Extremal K\"ahler metrics"}, Seminars
on differential geometry,  ed. S. T. Yau, {\sl Ann. Math. Stu.}, Princeton
University Press (1982), p. 259.

\bibitem{CValent-extremal}  T. Chave and G. Valent, {\sl Class. Quantum Grav.}
{\bf 13} (1996) 2097.

\bibitem{DS94Camb} A. S. Dancer and Ian A. B. Straham, {\sl Proc. Camb. Phil.
Soc.} {\bf 115} (1994) 513.

\bibitem{PD90} H. Pedersen and Y. S. Poon, {\sl Class. Quantum Grav.} {\bf 7}
(1990) 1707.


\bibitem{PDSRheine} H. Pedersen and A. Swann, {\sl J. reine angew. Math. } {\bf
441} (1993) 99.

\bibitem{Lebrun} C. Lebrun, {\sl Comm. Math. Phys.} {\bf 118} (1988) 591.

\bibitem{Gibbons79} G. W. Gibbons and C. N. Pope, {\sl Comm. Math. Phys.} {\bf
66} (1979) 267.

\bibitem{Pedersen85} H. Pedersen,  {\sl Class. Quantum Grav.} {\bf 2} (1985)
579.

\bibitem{GibHaw} G. W. Gibbons and S. W. Hawking, {\sl Comm. Math. Phys.} {\bf
66} (1979) 291.

\bibitem{Der83} A. Derdzinski, {\sl Comput. Math.} {\bf 49} (1983) 405.

\bibitem{Besse} A. L. Besse, {\sl ``Einstein Manifolds"}, Ergebnisse der
Mathematik und ihrer Grenzgebeite, 3. Folge, Band 10, Springer-Verlag Berlin
Heidelberg (1987).

\bibitem{Page} D. Page, {\sl Phys. Lett.} {\bf 79B} (1978)  235.

\end {thebibliography}

\end{document}